\documentclass[11pt,twoside]{article}

\usepackage{asp2004}
\usepackage{epsf}
\usepackage{psfig}
\usepackage{lscape}

\begin{document}
\title{Soft X-ray Spectroscopy of the Hot DA White Dwarf LB1919 and the PG1159 Star PG1520+525}  
\author{K. Werner,$^1$ J. J. Drake,$^2$ T. Rauch,$^1$ S. Schuh,$^3$ and A. Gautschy$^4$}  
\affil{$^1$Institut f\"ur Astronomie und Astrophysik, Universit\"at T\"ubingen, Sand~1, 72076 T\"ubingen, Germany\\
$^2$Harvard-Smithsonian Center for Astrophysics, MS 3, 60 Garden Street, Cambridge, MA 02138, USA\\
$^3$Institut f\"ur Astrophysik, Universit\"at G\"ottingen, Friedrich-Hund-Platz~1, 37077 G\"ottingen, Germany\\
$^4$Wetterchr\"uzstr. 8c, 4410 Liestal, Switzerland
}

\begin{abstract}
We have performed soft X-ray spectroscopy of two hot white dwarfs with
the \emph{Chandra} observatory using the \emph{Low Energy Transmission
Grating}.  The first target is the hot DA white dwarf LB1919 ($T_{\rm
eff}$=69\,000~K). This star is representative of a small group of hot
DAs whose metallicities lie well below predictions from radiative
levitation theory. The \emph{Chandra} spectrum shows a rich absorption
line spectrum which may allow to find the origin of the
low-metallicity nature of these DAs. The second target is
PG\,1520+525, a very hot non-pulsating PG1159 star. We find that it is
hotter ($T_{\rm eff}$=150\,000~K) than the pulsating prototype
PG\,1159$-$035 ($T_{\rm eff}=$140\,000~K) and conclude that both stars
confine the blue edge of the GW~Vir instability strip.
\end{abstract}

\section{The hot DA white dwarf LB1919}

The purity of WD atmospheres results from gravitational settling of heavy
elements. The lightest element, either hydrogen or helium, is floating on top of
the star, giving rise to the DA and non-DA spectral classes,
respectively. Hydrogen in hot DA atmospheres is almost completely ionized so
that the EUV/soft X-ray opacity is strongly reduced. As a consequence, DAs with
$T_{\rm eff}>30\,000$~K can emit detectable amounts of thermal soft X-ray
radiation, leaking out from deep, hot photospheric layers.

The \emph{ROSAT} all-sky survey, however, detected \emph{many} fewer white dwarfs
($<200$) than expected ($>5000$) -- X-ray emission turned out to be the exception
rather than the rule.  It was realized that additional absorbers must be present
in the atmospheres.  \emph{ROSAT} and \emph{EUVE} observations revealed that metals are the
origin of this additional opacity, and that the EUV spectral flux distribution
of hot DAs is most strongly determined by iron and nickel through their
enormously large number of absorption lines. Radiative
levitation can keep traces of these and other metals floating in the atmosphere
against gravitational pull (e.g. Chayer et al.\@ 1995): \emph{the metal
abundances therefore generally increase with increasing $T_{\rm eff}$.}  As a
consequence, only very few DAs with $T_{\rm eff}>60\,000$~K were detected in
the \emph{EUVE} and \emph{ROSAT} all-sky surveys because the increasing EUV opacity is
effectively blocking flux.

\begin{figure}[!t]
\begin{center}
\epsfxsize=\textwidth 
\epsffile{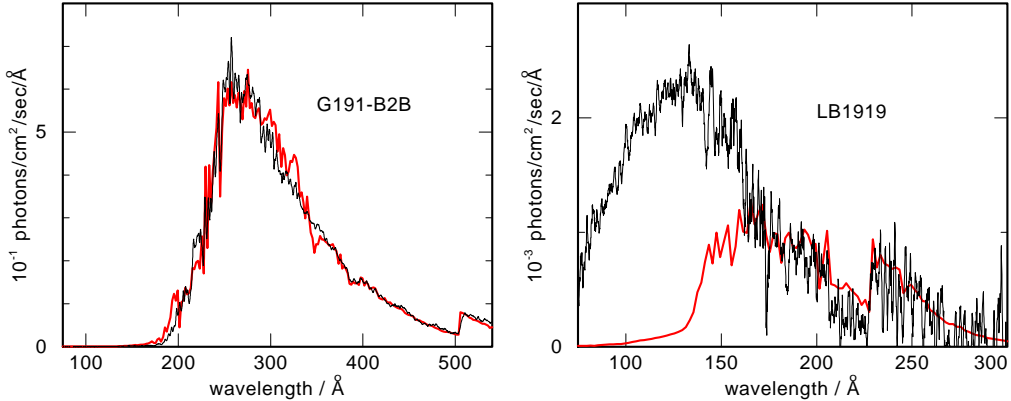}
\caption{\emph{EUVE} spectra of two hot DA white dwarfs
(thin lines) compared to spectra from models (thick lines) which
consider gravitational settling and radiative levitation. {\bf Left:}
The model fit to G191-B2B is very good. {\bf Right:} The observed flux
of LB1919 at $\lambda < 150$~\AA\ is much higher than predicted by the
model. For an unknown reason, the metallicity in LB1919 is peculiarly low.}\label{fig1_DAs.ps}
\end{center}
\end{figure}

A breakthrough in understanding WD atmospheres has been the successful
development of self-consistent models which describe the vertical stratification
of element abundances by assuming equilibrium between gravitation and radiative
acceleration.  These models no longer have the surface metal abundances as free
parameters: they are instead computed on physical grounds.  Generally, good
agreement is obtained between observed and computed EUV flux distributions
(Schuh et al.\@ 2002).  However, several exceptions are known.  Some DAs show
much larger metal abundances than expected from theory.  This has been explained
by either accretion from the ISM or wind-accretion from unseen companions.

Much more difficult to explain are those objects whose metallicity is
\emph{smaller} than expected from radiative levitation theory. One prominent
example is the famous white dwarf HZ43 ($T_{\rm eff}$=51\,000~K) whose
atmosphere is virtually metal free and shows no EUV or X-ray absorption
features.  Even more surprising is the low metallicity of DAs with even higher
temperatures. One of the hottest known DAs, LB1919 ($T_{\rm eff}$=69\,000~K),  has
extraordinarily low metal abundances.

LB1919 is the hottest DA of the 90 detected in the \emph{EUVE} all-sky survey (Vennes
et al.\@ 1997), and of the 20 DAs for which \emph{EUVE} spectra have been analysed in
detail by Wolff et al.\@ (1998).  The chemical composition is not known: since
the \emph{EUVE} spectral resolution is insufficient to unambiguously identify lines
from individual species, the metallicity of fainter DAs is usually determined
relative to that of the well-studied DA G191-B2B ($T_{\rm eff}$=56\,000~K).
Unlike LB1919, G191-B2B is sufficiently bright and metal-rich that its surface
composition can be constrained through UV spectroscopy.  Our stratified models
successfully describe in detail the EUV-UV spectrum of G191-B2B
(Fig.\,\ref{fig1_DAs.ps}).  The \emph{EUVE} spectra of other DAs could also be fitted
by simply scaling this relative metal abundance pattern.

\begin{figure}[!t]
\begin{center}
\epsfxsize=0.85\textwidth 
\epsffile{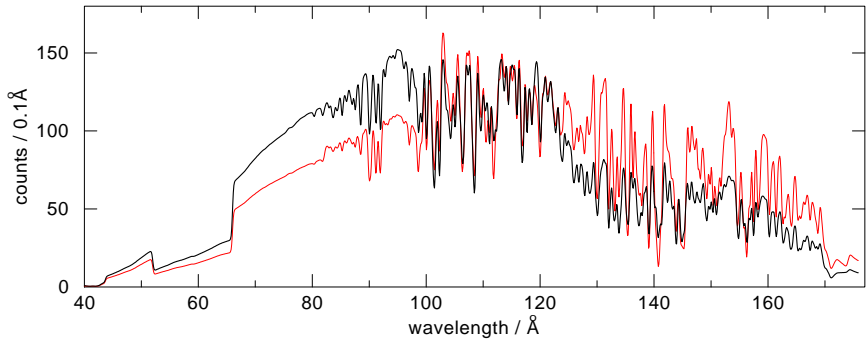}
\caption{
Simulated \emph{Chandra} spectra of LB1919 with homogeneously mixed models. 
Model parameters are $T_{\rm eff}$=70\,000~K, $\log g$\,=8.2. Following the
\emph{EUVE} analysis of Wolff et al.\@ (1998), the Fe
and Ni abundances are 10\% of the abundances determined for G191-B2B,
i.e., Fe/H$=7.5\cdot 10^{-7}$ and Ni/H$=5\cdot 10^{-8}$. In order to demonstrate
the sensitivity of the spectrum against abundance variations, we show two
models. The first has Ni increased by 1 dex (thick line), the second Fe
\emph{and} Ni increased by 1 dex (thin line).
}\label{fig3_DAs.ps}
\end{center}
\end{figure}

The \emph{EUVE} spectrum of LB1919, however, cannot be explained by
spectral models based on homogeneous abundances scaled relative to
G191-B2B, and stratified models including radiative levitation fail
spectacularly (Fig.\,\ref{fig1_DAs.ps}).  Its EUV spectrum suggests a
surprisingly low metallicity -- approximately 10\%\ of that of
G191-B2B, despite being significantly hotter.  The failure to match
\emph{EUVE} spectra is due to either the assumed G191-B2B-like
abundance pattern being wrong, or due to a different stratification
than predicted by pure radiative levitation.

Several processes
could disturb the equilibrium between gravitation and levitation and are
potentially responsible for the metal-poor hot DAs: mass-loss; accretion from
the ISM; convection, and mixing through rotation. Mass-loss would have the
effect of homogenising a chemical stratification, but it also has been shown
that mass-loss rates drop below the critical limit ($10^{-16}$~M$_\odot$/yr) for
DAs cooler than 70\,000~K so that this phenomenon should not occur in hot DAs
like LB1919 (Unglaub \& Bues 1998).  Wind accretion calculations show that
accretion is prevented for LB1919 since its luminosity is $>1$~L$_\odot$;
instead a mass-loss rate of $\approx 10^{-18}$~M$_\odot$/yr will be sustained
(MacDonald 1992).  Convection would homogenise abundances; however, the
atmosphere of LB1919 is
convectively stable because hydrogen is almost completely ionised. Rotation
could lead to mixing through meridional currents, but most WDs are very slow
rotators and analysis of a \emph{FUSE} archival spectrum of LB1919 shows deep and sharp
Lyman line cores which clearly excludes such a high rotation rate.  To
summarize, there is currently no explanation for the very low metallicity in
LB1919 and similar DAs.

\begin{figure}[!t]
\begin{center}
\epsfxsize=0.85\textwidth
\epsffile{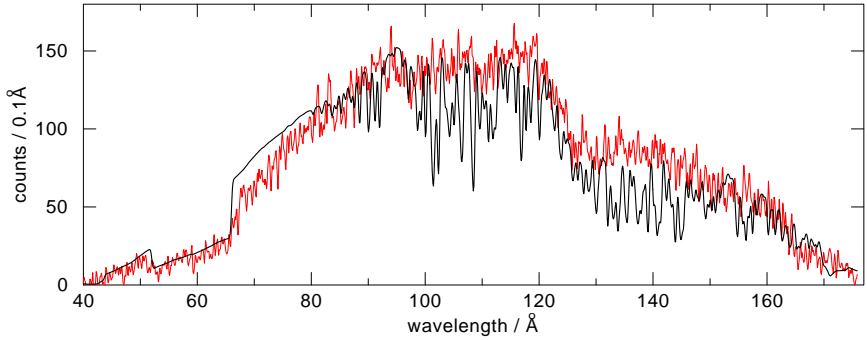}
\caption{\emph{Chandra} spectrum of 
  LB1919 (thin line). Overplotted is one of the model spectra depicted in
  Fig.\,\ref{fig3_DAs.ps} (Ni 10 times enhanced, Fe on the Wolff et al.\@
  value). The line features in the model are stronger than observed.
}\label{figlb1919}
\end{center}
\end{figure}

\begin{figure}[!t]
\begin{center}
\epsfxsize=8cm 
\epsffile{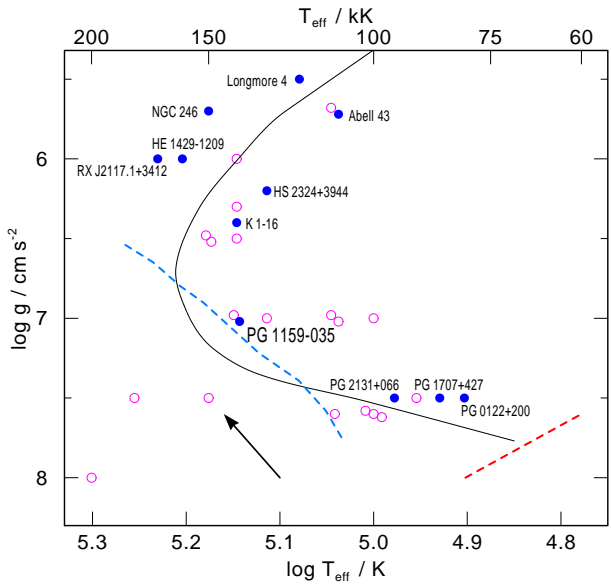}
\caption{The pulsating PG1159 stars (filled symbols with name-tags)
and the non-pulsators. The non-pulsator PG\,1520+525 is marked by an arrow. 
The dashed lines are theoretical blue and red
edges of the instability strip from Gautschy et al.\@ (2005) and
Quirion et al.\@ (2004), respectively. Also shown is a 0.6~M$_\odot$ post-AGB
track (full line) from Wood \& Faulkner (1986).}\label{figpulsators}
\end{center}
\end{figure}

\begin{figure}[!t]
\begin{center}
\epsfxsize=\textwidth 
\epsffile{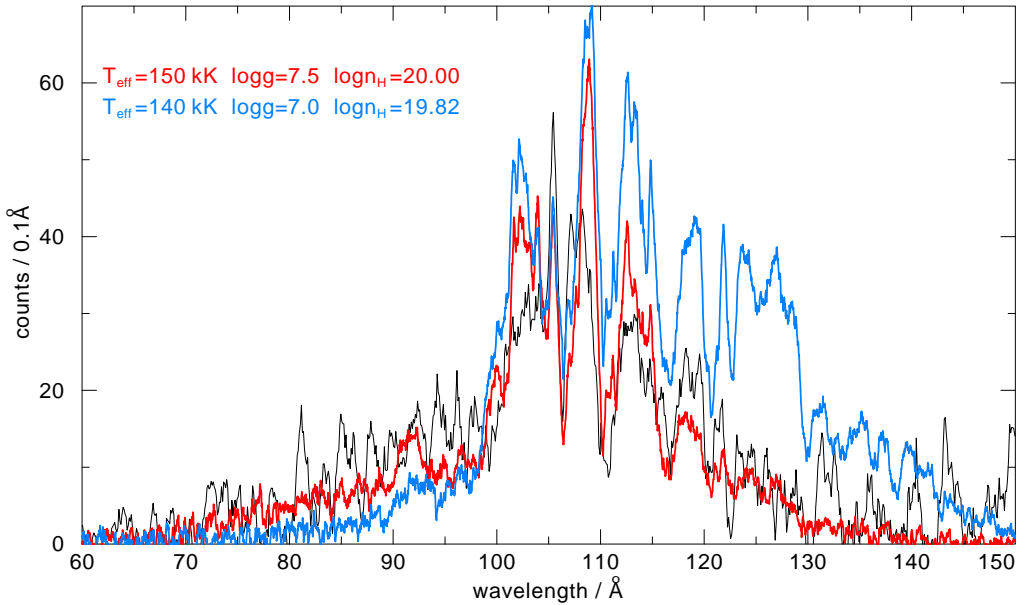}
\caption{\emph{Chandra} spectrum of PG\,1520+525
  (thin line), smoothed with a 0.5~\AA\ boxcar, and simulated
  observations from two models with different $T_{\rm eff}$. The
  140~kK model overestimates the flux at $\lambda > 110$~\AA. The
  models represent the pulsator--non-pulsator pair, see Tab.\,\ref{tab:pg1159stars} for
  atmospheric parameters.}\label{figcounts}
\end{center}
\end{figure}

In order to find out if the metals in the atmospheres of these peculiar
low-metallicity white dwarfs are stratified or homogeneous we decided to perform
\emph{Chandra} spectroscopy of LB1919 with the \emph{Low Energy Transmission
Grating} (LETG/HRC-S). It was demonstrated by Vennes et al.\@ (2002) that individual lines
can in principle be identified in \emph{Chandra} spectra of hot DAs (GD246).
A simulation with our model spectra shows that we should also be able to identify
individual lines in LB1919 (Fig.\,\ref{fig3_DAs.ps}). In Fig.\,\ref{figlb1919}
we show the observed spectrum. It was taken on Feb.~02--03, 2006, with an
integration time of 111~ksec, and it clearly shows many spectral lines. The
overplotted spectrum of our preliminary model reproduces the
spectral shape, however, the absorption lines are obviously too strong.

\section{The PG1159 star PG\,1520+525}

PG1159 stars are hot H-deficient (pre-) WDs which are probably the outcome of a
late He-shell flash (Werner \& Herwig 2006). The pulsating members of this
spectral class form the GW~Vir instability strip in the HRD. In order to
constrain stellar evolutionary models it is of interest to empirically find the
edges of this strip. The pulsating prototype PG\,1159$-$035 and the non-pulsator
PG\,1520+525 are located close to each other in the $g$--$T_{\rm eff}$ plane
(Fig.\,\ref{figpulsators}) and it can be argued that the blue edge of the strip
is constrained by this pair of stars. From optical and UV spectroscopy it was
found that PG\,1520+525 is indeed hotter than PG\,1159$-$035
($150\,000\pm15\,000$~K and $140\,000\pm5\,000$~K, respectively), however, the
error bars of the analyses overlap considerably. While we could constrain the
temperature of  PG\,1159$-$035 with high precision (Jahn et al.\@ 2007), the
uncertainty for PG\,1520+525 is still very large (Dreizler \& Heber 1998).

\begin{figure}[!t]
\begin{center}
\epsfxsize=\textwidth 
\epsffile{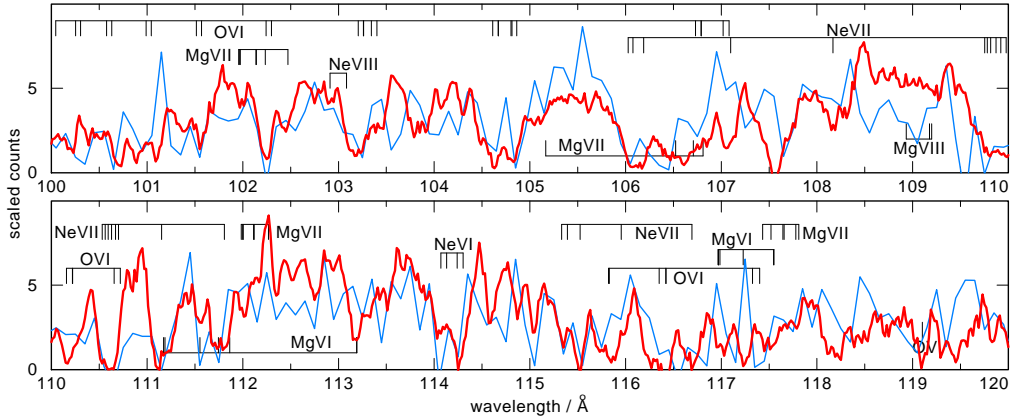}
\caption{Detail of the \emph{Chandra} spectrum of PG\,1520+525
  (thin graph) and our 150~kK simulation. Lines from
  \ion{O}{vi} and \ion{Ne}{vi/vii} can be identified.}\label{figchandra_fit_detail}
\end{center}
\end{figure}

We have taken a \emph{Chandra} LETG/HRC-S spectrum because our models predict
a strong sensitivity of the soft-X-ray flux to variations in $T_{\rm
eff}$. The observation was performed on April 04--06, 2006, with an
integration time of 142~ksec. We show the spectrum in Fig.\,\ref{figcounts}
together with two models of different temperature. There is evidence that the
hotter model fits much better and we hope that a detailed analysis will give
$T_{\rm eff}$ with a precision of $\pm 5000$~K. This analysis requires to fit
individual features in the spectrum. Fig.\,\ref{figchandra_fit_detail} shows
that we can identify a number of absorption lines from highly ionised oxygen and
neon. At present it seems that we can confirm that PG\,1520+525 and PG\,1159$-$035
do indeed constrain the blue edge of the GW~Vir instability strip. The
interpretation of our results, however, must consider that the
exact position of the edge is depending on the photospheric composition (Quirion
et al., these proceedings).

\begin{table}[!t]
\begin{center}
\vspace{-0.5cm}
\caption{Atmospheric parameters of the pulsator--non-pulsator pair. Abundances
  are given in mass fractions.}\label{tab:pg1159stars} 
\footnotesize
\begin{tabular}{lcc} 
\noalign{\smallskip}
 \hline
\noalign{\smallskip}
            & PG\,1159$-$035 & PG\,1520+525 \\
\hline
\noalign{\smallskip}
$T_{\rm eff}$/K & 140\,000 & 150\,000 \\
$\log g$      & 7.0      & 7.5 \\
He            & 0.33     & 0.43 \\
C             & 0.48     & 0.38 \\
O             & 0.17     & 0.17 \\
Ne            & 0.02     & 0.02 \\
$\log$ F      & $-5.5$     & $-4.0$ \\
$\log$ N      & $-3.0$     & $< -3.0$\\
\hline
\vspace{-1cm}
\end{tabular} 
\end{center}
\end{table}

\acknowledgements T.R.\@ was supported by DLR grant 50 OR 0201. J.J.D. was supported by NASA contract NAS8-39073 to the {\it Chandra
X-ray Center}.

\end{document}